\documentclass[prl,amsmath,amssymb,showpacs,showkeys,twocolumn]{revtex4}
\usepackage{graphicx}
\usepackage{dcolumn}
\usepackage{bm}
\usepackage{amssymb}
\usepackage{amsmath}
\begin{document}

\title{Magnonics and Supermagnonics}

\author{Yury Bunkov}

\address{Russian Quantum Center,  Skolkovo, Moscow, 121205, Russia.\\
\email{yury.bunkov@neel.cnrs.fr}
}

\begin{abstract}

The magnetic community continues to discuss the possibility to observe the magnetic superfluidity, despite the fact that it has been discovered long time ago. It was observed in antiferromagnetic states of superfluid $^3$He in 1984.
In this article we  reminds the main principles of spin superfluidity and related Bose-Einstain magnon condensation. We discuss applications of this phenomenon in supermagnonic devises.

\end{abstract}


\maketitle

\section{Spin current and spin supercurrent}

The magnetically ordered systems characterized by an order parameter, which is the result of spontaneous breaking of spin-orbit $SU(2)$ symmetry in the equilibrium magnetic states. These states shows the rigidity with respect to the inhomogeneous
 spin rotations $\theta_\alpha({\bf r})$ i.e. to the dependence of the energy on the gradients $\omega_{\alpha i}=\nabla_i \theta_\alpha$. This rigidity leads to formation of spin waves (the collective oscillations of magnons) and spin transport in magnetic textures.  Particularly the flow of magnetization can circulate in the case  of  topological defect in magnetically ordered materials. The spatial spin current can be excited also by pumping of magnetization in one side of the sample and its sink in the other side. The experimental observation of this current is discussed for a long time. See \cite{A,B,C} and references there. The spin current in this case is proportional to the gradient of order parameter. Indeed its nature  is very different from an origin of superfluid and supercurrent phenomena. Let as consider the analogy with the cosmology. The ground state of the Universe - the quantum vacuum is the ordered state described by a complex matrix. The gradients of this state leads to a flow of vacuum, particularly in the case of cosmic strings. Similar case take place in magnetically ordered states. The gradient of  ground state leads to a flow of magnetization. Can we call this flow a supercurrent? Yes and Not. From one side the gradient of ground state leads to the flow of magnetization. But originally the term superfluidity and superconductivity applied for the flow of excitations and not a flow due to the gradient of ground state - quantum vacuum. 

This difference is very clear in magnetically ordered systems. There is a ground state and excitations - magnons. Physicist usually deal with the dynamics of magnetic ground state at a small excitation described by Landau-Lifshits equations. But the situation drastically changes in the case of a big density of magnons. In this case magnons may form a Bose-Einstein condensate, which gradients leads to a spin supercurrent in full analogy with mass superfluidity and superconductivity. This type of flow has all rights to be called supercurrent while the flow of magnetization in the case of magnetic texture of ground state we can call textural current.

 Since magnons obey the Bose statistics, they may form Bose-Einstain condensate (BEC) similar to an  atomic BEC.  Atoms forms a coherent quantum state at the temperature below a critical one for given density of atoms. These conditions was predicted by Einstein \cite{Einstein} and follows from Bose statistics:
\begin{equation}
T_{BEC}\, \simeq 3.31\,\frac{\hbar ^{2}}{k_{B}m}\left({N_C}\right)
^{2/3}, \label{TBECpart}
\end{equation}
where $N_C$ is the density of atomic gas and $T_{BEC}$ is the critical temperature, below which atoms condense in a BEC state. 
The magnon BEC state should forms at about the same ratio between the temperature and density. Magnons condensed to a Bose-Einstein condensate
at a higher density. 

Indeed, magnons have a finite lifetime and the total number of equilibrium magnons decreases with decreasing of temperature (and reach zero at $T=0$).
The density of magnons at thermodynamic equilibrium is always below the critical density of  magnons BEC formation.
However, the density of excited non-equilibrium  magnons $N_M$ can be drastically increased up to about Avogadro density by a magnetic resonance methods.
The excited magnons can form the quasi-equilibrium excited state with a time scale of about few quasiparticles scattering time.
The usual 4-magnon scattering conserves the total number of quasiparticles and hold the distribution function with the effective temperature $T$ and effective chemical potential $\mu$.
The critical density of magnons BEC formation $N_{BEC}$ can be estimated from equation (1). The initial temperature of magnons correspond to  phonon subsystem temperature, which determines the value of magnetization $M$ and density of thermal magnons.  We are able to increase the density of excited magnons above the critical one by dynamical magnetization deflection. Magnons should forms a BEC state when $N_M > N_{BEC}$, under certain conditions, which will be discussed below. The critical magnons concentration $N_{BEC}$ for ferro and antiferromagnets was calculated in \cite{critdens,Bunkov2018a}. Particularly for easy plain antiferromagnets with wave spectrum
\begin{equation}
\varepsilon_{k}=\sqrt{\varepsilon _{0}^{2}+\varepsilon _{ex}^{2}(ak)^{2}}
\label{AF spectrum}
\end{equation}
it reads:
\begin{equation}
N_{BEC}\simeq \frac{(k_{B}T)^{2}}{2\pi ^{2}}\frac{%
\varepsilon _{0}}{a^{3}\varepsilon _{ex}^{3}}.
\label{NBECpart}
\end{equation}

The magnetic oredering in a magnon BEC state arises due to spontaneous breaking of $U(1)$ symmetry in the non-equilibrium magnon condensate. In atomic BEC and in helium superfluids such symmetry breaking leads to a non-zero value of the  superfluid rigidity --  the superfluid density  $\rho_s$ which enters the non-dissipative  supercurrent  of particles and thus to mass supercurrent. The same takes place for magnon BEC. 
The supercurrent of magnons can be described by traditional equations:
\begin{equation}
{\bf J}= \rho_s {\bf v}_s~~,~~{\bf v}_s =\frac{\hbar}{m_M}{\bf \nabla}\alpha~~,~~\rho_s(T=0)=N_M m_M ~.
 \label{MassCurrent}
 \end{equation}
where $m_M$ is the effective mass of magnons.
 In translationally invariant systems, where the mass current coincides with density of linear momentum,  Eq.(\ref{MassCurrent}) can be obtained directly from the definition of linear momentum density in spin systems:
\begin{equation}
{\bf P}=(S-S_z){\bf \nabla}\alpha=  N_M \hbar {\bf \nabla}\alpha\,,
\label{MassCurrentMagnon}
\end{equation}
where we used the fact that dencity of magnons $N_M = S-S_z$ and phase of magnetization precession $\alpha$ are canonically
conjugated variables. As in conventional superfluids, the superfluid density of the magnon liquid is determined by the magnon density $N_M$ and magnon mass $m_M$.
The superfluid mass current (\ref{MassCurrent}) carries magnons with  mass $m_M$, while in atomic superfluids  the superfluid mass current carries atoms. The magnon mass current generated by precessing
 magnetization in magnon BEC is similar to electric current generated by  precessing
 magnetization in ferromagnets \cite{Volovik1987}. 

The best magnetically ordered spin systems for magon BEC and spin supercurrent investigations are antiferromagnetic states of superfluid $^3$He owing the very small Gilbert damping (about 10$^{-9}$), its absolute purity and different types of magnon-magnon interactions. It is very important to know that the dynamic properties of these states are the results of magnetic ordering and does not related with its mass superfluid properties \cite{MagBEC}.

\section{Spin superfluidity}

In 1984 the first magnon BEC state and related spin superfluidity  has been discovered in antiferromagnetic superfluid $^3$He. The magnon BEC demonstrate the spontaneously self-organized phase-coherent precession of spins \cite{HPD,HPDT}. This state is radically different from the conventional ordered states in magnets. It is the quasi-equilibrium state, which emerges on the background of the ordered magnetic state, and which can be represented in terms of the Bose condensation of magnetic excitations -- magnons \cite{MagBEC}. 

The magnon BEC opened the new class of the systems, the Bose-Einstein  condensates of quasiparticles, whose number is not conserved. Representatives of this class in addition to BEC of magnons are the BEC of
phonons \cite{2}, excitons \cite{3}, exciton-polaritons
\cite{4}, photons \cite{5}, rotons \cite{6} and other  bosonic quasiparticles.

Owing the coherence the magnon BEC radiates a very Long Living Induction Decay Signal (LLIDS). It may be considered as a time crystals \cite{Tcrystal}  with a  very long, but finite lifetime. It may reach minutes in antiferromagnetic superfluid $^3$He-B. Furthermore, the Goldstone modes - the time-space excitations of the time crystal (the analog of second sound in superfluid $^4$He)  have been observed in magnon BECs \cite{Goldstoun1,Goldstoun2,Goldstoun3}. The lifetime of magnon BEC states may be infinite in the case, when the losses (evaporation) of quasiparticles are replenished by an excitation of new quasiparticles. 

The formation of a magnon BEC state was first observed in antiferromagnetic superfluid $^3$He-B 
\cite{HPD,JETPBEC}. Usually, in the linear case,  the induction decay signal after a RF pulse ringing the time, inversely proportional to an inhomogeneous broadening of magnetic system $\Delta \omega$. In the experiments, described in \cite{HPD,JETPBEC} the induction signal also lost coherency at the time scale about $ 1/ \Delta \omega$ but then spontaneously reappears and ringing a few orders of magnitude longer!
The formation of LLIDS manifest itself the condensation of magnons in a common wave function in all the sample with a common phase and frequency of precession. 
The mechanism of BEC state formation related to a repulsive intersection between magnons. The higher local magnon density - the bigger dynamical frequency shift in the system. The local inhomogeneity of Larmore precession generates the gradient of phase of precession and, consequently, the superfluid transport of magnons. This spin supercurrent redistribute magnons until the dynamic frequency shift compensate the magnetic field inhomogeneity. The so named Homogeneously Precessing Domain (HPD) forms. The of coherently precession magnetization in HPD radiates LLIDS signal.

 The LLIDS  obey all the requirement for BEC of quasiparticles, which much later was postulated as an requirement of magnon BEC in well known article by Snoke \cite{Snoke}. Magnon BEC has one to one analogy with the experiments of atomic  BEC \cite{mBEC}. Owing the slow magnons relaxation, the number of magnons decrease, but the magnons remains in a coherent state. It is important to note that the BEC state is the eigen state of excited magnons. It was shown experimentally, that the small RF pumping on a frequency of magnon BEC $\omega_{BEC}$ can compensate the magnons relaxation by creation of an additional  magnons. In this case the magnons BEC may maintains permanently for an infinite time \cite{CWBEC}.   The 35 years of magnons BEC investigations in different antiferromagnetic states of superfluid $^3$He well established the physics of excited magnon BEC and phenomena of spin superfluidity.
The review of this investigations one can found, for example in \cite{Rev1,Rev2} and in the book \cite{MagBEC}.

\section{Experimental observation of spin supercurrent}

The excitation of the homogeneously precessing domain is an interesting
discovery in itself. The next step in investigations of the magnon BEC  was the experimental studies of spin supercurrent between two independent HPD states, connected by a channel which was either perpendicular  \cite{6,62,R1} or parallel to magnetic field \cite{63}.
The idea of these experiments was very straightforward and based on the
analogy of a superconducting bridge between two massive superconducting
electrodes. Here we can consider two cells filled with HPD as such
electrodes.  The
role of the  potential difference between the electrodes is
equivalent to the difference of the HPD precession frequencies. This
difference leads to an increase in the gradient of phase
of precession
in the channel and consequently to the growth of spin supercurrent.  If
one keeps the frequencies of HPD's precession the same, then the
phase gradient in the channel remains constant and a steady state
supercurrent has to pass through the channel.  In the case of
superconductivity this current is supplied by the leads of the normal
metal, which have some resistance and consequently there is a voltage difference.
In the case of the spin supercurrent the longitudinal magnetization is not
conserved in the RF field. Therefore, the RF field can pump the
longitudinal magnetization into one cell and pump it out in the other
cell.  The transport of the longitudinal magnetization along the
channel in a magnetic field is accompanied by transport of  Zeeman
energy. This transport has been measured by the
increase in one cell of the energy
absorbed from the RF field with increasing phase difference, and its decrease in the other cell.  In this
way we were able to measure the current of longitudinal magnetization
flowing out from one cell and  into the other.

  Owing to the direct relation between the phase of magnetization precession
and the phase of the order parameter,  we were able to control the spin supercurrent  through
 measurement of the phase gradient of the magnetization precession in the channel.
This method has no
analogue with superconductivity because there is no field that is
sensitive to the phase of the wave function of electron Cooper pairs.

\begin{figure*}[htt]
\begin{center}
 \includegraphics[width=3.5in]{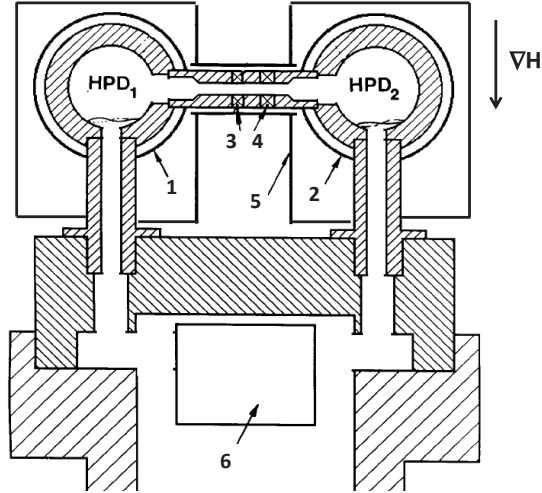}
\end{center}
 \caption{The experimental cell for studies of spin supercurrent between two HPD states, excited by RF from two independent coils 1 and 2. The channel of 1.4 mm in diameter connecting  HPD states. The pick up coils 3 and 4 monitoring the amplitude and phase of the magnetization  precession inside the channel. Pt NMR thermometer (6) monitoring the temperature. The screen (5) suppress the RF crossover signal betwen the coils. The cells connected to a main volume of $^3$He by a channels. }
 \label{Chanal}
\end{figure*}

\begin{figure*}[htt]
\begin{center}
 \includegraphics[width=3.5in]{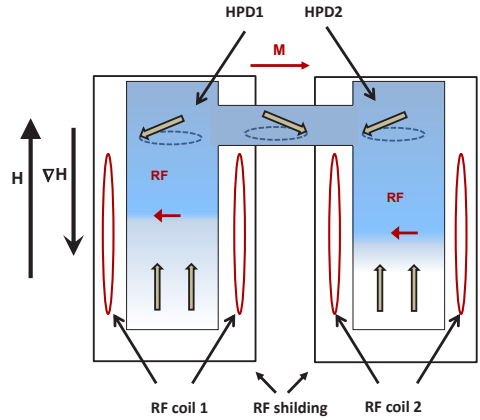}
\end{center}
 \caption{Illustration of experimental observation of spin supercurrent between two HPD states. The spin supercurrent transport the magnetization from cell 1 to cell 2. The current is proportional to the gradient of phase of precession in the channel. }
 \label{Chanal}
\end{figure*}
The first observation of a spin supercurrent in
the channel have been published in \cite{6}.
The experimental set-up  consists of two cells in
the form of a barrel with axes parallel or perpendicular to the magnetic
field, connected by a channel perpendicular to field (see Fig. \ref{Chanal}).
 The cells were
surrounded by RF coils, and copper shielding prevented
interaction between the coils. The channel was surrounded by
additional shielding to prevent RF field penetration into the
channel.  The coils 1 and 2 were used to excite  HPD states
in both cells and to control them.  The frequency and phase of the
precession of the domain with homogeneous precession in each of the
volumes was determined by the frequency and phase of the radio-frequency
field of the corresponding coil,  supplied from separate highly stable
generators. The cells were filled  with HPD by sweeping down the
magnetic field. When the domain boundary crossed the inlet to the
channel, the HPD filled the channel.  Miniature receiving
radio-frequency coils 3 and 4 were set up in the channel, and
received a signal from the precessing magnetization in the channel.  A
small signal induced by the exciting coils was compensated by an
electronic circuit.  For HPD creation, equal frequency and phase of both RF
generators was chosen, so we can assume that the difference of phase of
precession in the channel is zero. Then the frequency of one of the rf
generators was changed by $\delta \omega \simeq 0.1$Hz. This causes
the difference between phases of precession
$\Delta \alpha$ to grow. A phase difference between  two HPD's determines the 
phase gradient along the channel.

\begin{figure*}[htt]
\begin{center}
 \includegraphics[width=4in]{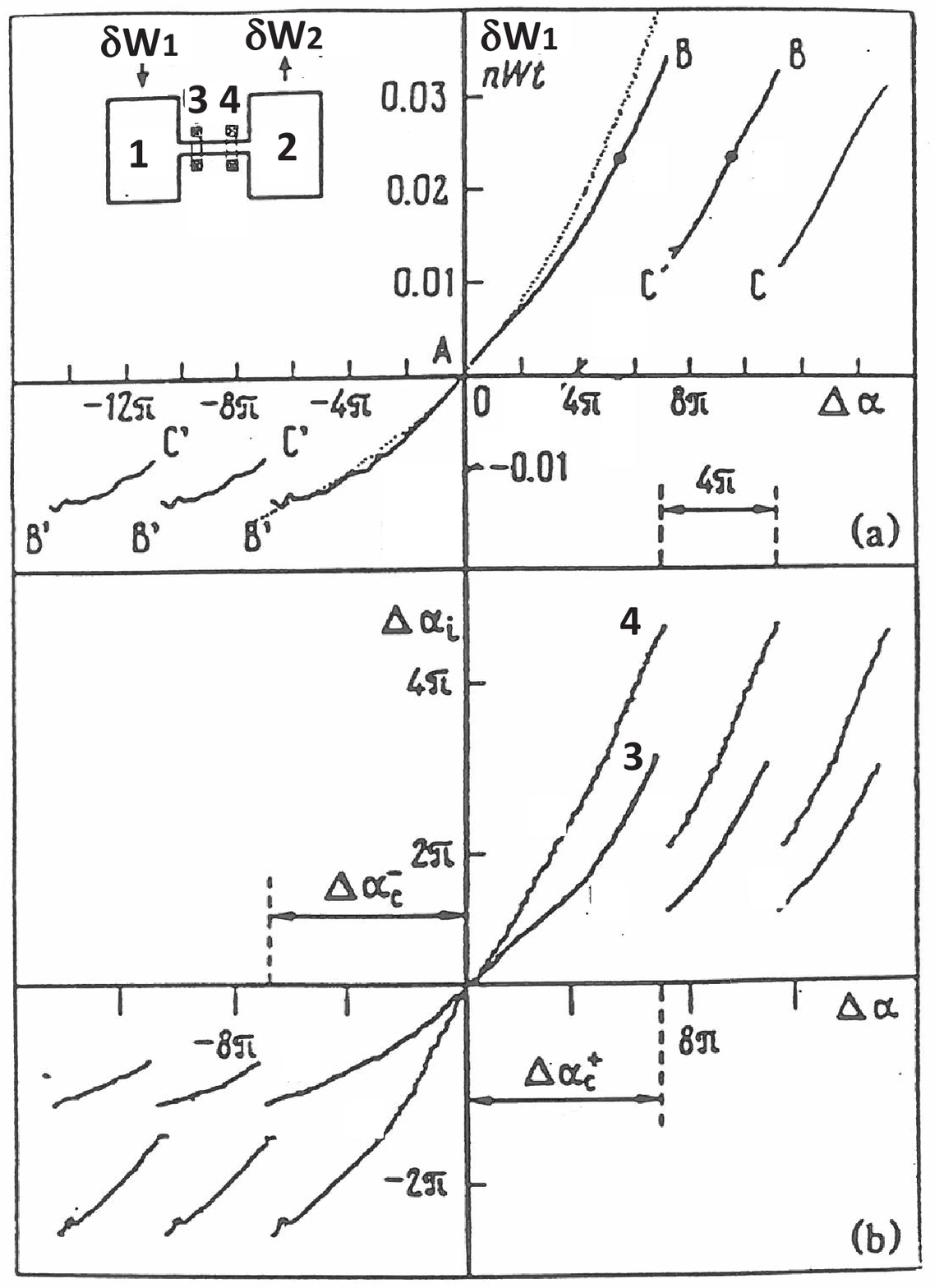}
\end{center}
 \caption{The record of energy dissipation in both HPD's as function of phases of prenessin in HPD's and the phase of precession in pick-up coils 3 and 4. The spin supercurrent transport the magnetization from cell 1 to cell 2. The absorption of energy in cell 1 increase for to compensate the zeeman energy, transported out by spin supercurrent. This energy partly dissipate in the channel due to diffusion relaxation and partly arrive to cell 2. Consequently the energy, absorbed from the RF field in cell 2 decreasing. At point B the phase slippage of $4\pi$ appears. This critical spin current in the channel determins by  $\omega-\omega_L$ and thus as a function
of the coherence length $\xi$ in  (\ref{Ksi}). The signals from pick-up coils measured the phase distribution inside the channel.}
 \label{2cell}
\end{figure*}

If we keep increasing this phase difference, the spins in the channel will
``wind up" to maintain boundary conditions. The spin current in the
channel increases, until it reaches a certain critical point, after which it drops
by a certain, specified amount. Here, the misequationment of the spins is too
great to warrant transfer, and locally the HPD is disrupted, and the spins in
the channel ``unwind''. At this spot, the magnetisation will locally be equationed
parallel to the external magnetic field, and in this way ``lose'' several times
$2\pi$ of twisting. Once the tenseness has gone out of the system, the spins
reequation with the surrounding precession angles, and the spiral reforms with
a few windings less. If the precession phase difference is increased again,
they will wind up until the critical value once more, as is shown by the
measurement in Fig. \ref{2cell}   There is shown the rise of the absorption
signal in cell 1  and its diminishes in cell 2. ( Due to
the symmetry, the signal from cell 1 at negative  $\Delta \alpha $
corresponds to the signal from cell 2 at positive $\Delta \alpha $).
This process corresponds to a transfer of  longitudinal
magnetization, and consequently the Zeeman energy,  from one chamber to the
other.  
All experimental curves correspond to stationary
solutions in the channel. To check this, we made the frequencies
 of the HPD's equal at a certain time. Then the
 absorption signals from both HPD and gradient distribution in the
 channel did not change any more - a steady state spin supercurrent
 continued to flow along the channel.

With increasing $\Delta\alpha $  one can see that on reaching a critical
phase difference $\Delta \alpha _c^+$ at point B the absorption jumps
to a smaller value (point C), then increases to the critical value
again, etc. In this case the  jumps occur with period $2n\pi $ in $\Delta\alpha $.
The critical  phase gradient determines by the  inverse
value of the
Ginzburg-Landau coherence length.

\begin{equation}
\nabla \alpha^2 _c=1/\xi^2_{GL}=\omega_L (\omega_{RF}-\omega _L)
/c_\perp^2.
\label{Ksi}
\end{equation}

Its increase with the increasing the difference $\omega_{RF}-\omega _L$. The value of phase slippage also increase. Experimentally the phase slip up to  $16\pi $ have been observed. Similar
jumps can be seen in the phase of precession in the channel.
The gradient of the phase of
precession in the channel produces a spin supercurrent which, for the
channel perpendicular to H, reads:

\begin{equation} 
J_P=-{\chi \over \gamma }(1-\cos \beta)[(1-\cos
\beta)c_\parallel^2+(1+\cos \beta)c_\perp^2] \nabla \alpha. \label{9.1}
\end{equation}

This supercurrent transports the longitudinal magnetization from cell 2
to cell 1. The rise of the magnetization in cell 1 means a decrease of
the angle $\beta$. To maintain the resonance condition, the HPD in this
cell begins to absorb more RF power (curve AB). The same supercurrent
leads to an increase of angle $\beta$ in cell 2. To prevent this the NMR
absorption must fall in this cell (curve AB'). In other words the
magnetic supercurrent transports some magnetic energy $J_E=-J_PH$ from
cell 1 to cell 2. To compensate this energy flow the rf absorption
rises in one cell by $\delta W_1$ and falls in the other one by
$-\delta W_2$.  If the magnetization transported by the supercurrent
were conserved, we would have $\delta W_1 = - \delta W_2$. However,
there are some relaxation processes caused by interaction between the
magnetization of the normal and superfluid components. Spin diffusion
of the normal component leads to a dissipation of magnetic energy in
the channel, that grows with phase gradient. To maintain the resonance
conditions for the HPD in the channel, the energy losses should be
compensated by additional energy supply by spin supercurrent. So the
spin current is greater at the inlet of the channel than at the
outlet. The asymmetry of the experimental curve  about
$\Delta \alpha =0$ is the result of magnetic relaxation within the
channel. But this relaxation is not the result of friction, it can be
treated as a relaxation of the eigenstate, which can not be seen in the
case of mass superfluidity or superconductivity due to the conservation of
mass and charge.  By taking this relaxation into account one can
recalculate the distribution of $\nabla \alpha $ along the channel:

\begin{equation}\nabla \alpha (x)={\exp{\Lambda \Delta \alpha }-1 \over
\Lambda [L+(\exp{\Lambda \Delta \alpha }-1)x]} \label{9.2}
\end{equation}
where $L$ is the length of the channel and $\Lambda={64 \over
145}D\omega _{RF}c_\perp^{-2}$, where $D = (D_\parallel +D_\perp)/2$ is
the effective spin diffusion coefficient along the channel as
defined in \cite{Einzel}. With cooling this relaxation significantly decrease due to decrease of the normal component of liquid. Indeed, we was not able to investigate this phenomena at temperature below 0.4 T$_C$ due to instability of HPD state, which was named ``Catastrophic relaxation'' \cite{Cat}.

\begin{figure*}[htt]
\begin{center}
 \includegraphics[width=3.5in]{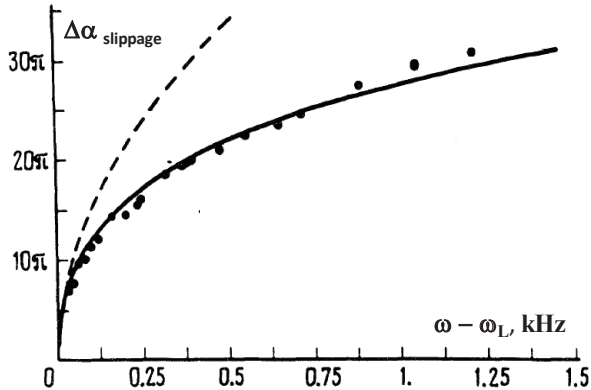}
\end{center}
 \caption{Critical phase difference versus frequency shift in the channel, measured at 29.3 bar and
1.4 mK. Dashed curve represents the theory for nonrelaxing magnetization, while solid curve is the
theoretical fit with spin-diffusion relaxation as a parameter. }
 \label{slippage}
\end{figure*}

\section
{ Phase Slippage.}

  The spin supercurrent in a channel is limited by the instability of
current against  phase slippage. In this section we shall analyse
the nature of phase slip centres for spin supercurrent. From a
general point of view the phase slippage of spin supercurrent is
analogous to that
observed in superconducting wires \cite{Kopnin} and mass superflow through a small hole \cite{Varoquaux}.
 We have learned from these superfluidity and
superconductivity experiments that the superfluid density should be
zero at the phase slip centre. As a result the phase of the
order parameter is not determined and the phase relation along
the channel can have a discontinuity. The formation of phase slip
 is related to a change in some energy. If this energy is less, than the
density of the
kinetic (gradient) energy of the supercurrent, the phase slip
appears. As a result of  phase slippage the phase difference  along
a channel  will be decries on $2n\pi$.  Upon decreasing the  kinetic (gradient) energy density
the phase slip centre becomes unstable and
disappears. The main difference between  phase slip in superfluidity and
superconductivity  and the phase slip
 of spin supercurrent  is that in the latter case it is not necessary to
destroy the superfluid state  to create the phase slip.
 It is
sufficient to destroy the spin supercurrent density which is
proportional to $(1-\cos \beta)$ (see (5.1))
 to maintain the spin supercurrent phase slip centre.
If $\beta =0$ in any part
of the channel,  the phases of
precession of the HPD in the cells are no longer connected
and the phase difference between the two HPD's can change by a multiple of
$2\pi $.  The critical current for creation of the
phase slip can be estimated by comparing the stiffness
of the HPD state in a channel and kinetic (gradient) energy of a current.
This corresponds  to the phase gradient equal to the inverse
value of the
Ginzburg-Landau coherence length according to Eq.(\ref{Ksi})

 As was shown in \cite{Fominslip}  the local
gradient energy is equal to the energy of HPD formation.
In reality the situation is more complicated. One should take
into account the spectroscopic correction to
the gradient energy that leads
to the frequency shift of
precession $ \Omega_\nabla= \partial F_\nabla /\partial P$ :

\begin{equation} \Omega_\nabla = {5c_\parallel^2-c_\perp^2 \over 4 \omega
}\nabla \alpha ^2 \label{5.3}
\end{equation}

The value of the
dipole-dipole frequency shift  decreases with increasing current  in order to
compensate this gradient energy frequency shift and
to keep the HPD in the channel in resonance. But when $
\Omega_\nabla $ surpasses the difference between the HPD frequency and the Larmor
frequency in the channel, the HPD can no longer exist and the angle
$\beta$  decreases.  Therefore the density of $P$,
proportional to $(1-\cos\beta)$, decreases
which makes the spin current
solution unstable.
Interestingly an analogous instability takes place in the case of
mass supercurrent in  $^3$He-B due to Fermi liquid corrections. As
was shown in \cite{VollhardtWolfle1990},
the superfluid density in  $^3$He-B decreases
with increasing  gradient of phase of the wave function (velocity).
Consequently the critical supercurrent corresponds to a maximum
value of current as a function of this gradient.
By taking into account the circumstances given above, the critical spin
supercurrent should correspond to the gradient:

\begin{equation}\nabla\alpha _c= \sqrt{4 \omega_L (\omega _{RF}-\omega
_L) \over 5c_\parallel^2-c_\perp^2 } \label{9.4}
\end{equation}

In Fig. \ref{slippage} we show the experimental value of the critical phase
difference between two HPD as function of $\omega _{RF}-\omega
_L$. In order to compare these results with theory one should take into
account the distribution of phase gradient in a channel, given by
\ref{9.2}.
There is a good agreement with the theory, particularly if we use
the spin diffusion coefficient as a fitting parameter, that is $
D=0.035\; cm^2/s$. (solid line in Fig. \ref{slippage}).  For the $D_\perp$,
measured under the same conditions we have $D_\perp =0.058\; cm^2/s$.  This discrepancy is probably
caused by spin - diffusion anisotropy \cite{Einzel,Sauls}, demonstrated
experimentally for the first time.

\begin{figure*}[htt]
	\begin{center}
		\includegraphics[width=3.5in]{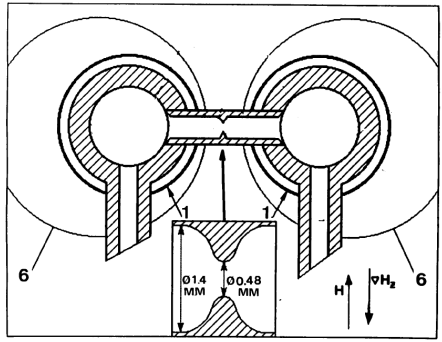}
	\end{center}
	\caption{ For observation of the dc and ac Josephson effects the orifice of diameter about 0.48 mm was installed inside the channel. 
	}
	\label{oreface}
\end{figure*}

\begin{figure*}[htt]
	\begin{center}
		\includegraphics[width=3.5in]{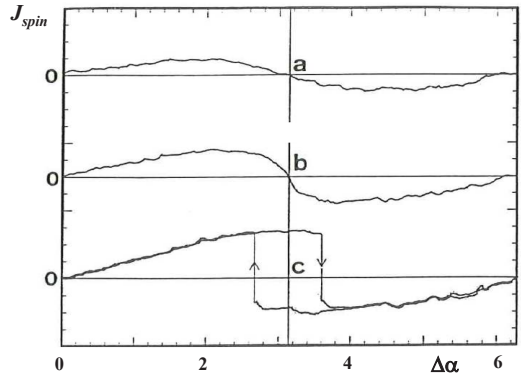}
	\end{center}
	\caption{  The  Josephson effect for magnon BEC  demonstrates the interference between two magnon condensates. Spin current as a function of the phase difference across the junction,
		$\alpha_2-\alpha_1$, where $\alpha_1$ and  $\alpha_2$ are phases of precession in two coherently precessing domains.
		Different experimental records correspond to a different ratio between the diameter of the orifice and the coherence length $\xi$ of magnon BEC.  The pure dc Josephson phenomenon was observed for magnetic coherent length $\xi=1.3$ mm (a)  and the distorted one for $\xi=0.8$ mm (b). The phase slippage processes were observed for $\xi=0.7$ mm (c).
	}
	\label{Jos}
\end{figure*}

\section{Spin-current Josephson effect}
\label{SpinJosephson}

The Josephson effect is the response of the current to the phase between two
weakly connected regions of coherent quantum states. It was described by Josephson
\cite{Joseph}  for the case of two quantum states, separated by the potential barrier.
This phenomenon is usually studied for the case of quantum states connected
by a conducting bridge with the dimensions smaller than the coherence
length. In this case the coherent state in the bridge cannot be established so there
is no phase memory, which determines the direction of the phase gradient. As a
result the supercurrent is determined only by the phase difference between the
two states.
As the dimensions of the conducting bridge increase, the more complex   current-phase
relation is observed. For bridge dimensions of the order of the
coherence length, a transition to a hysteretic scenario with phase slippage appears.

In the case of mass and electronic supercurrents the coherence length is a
function of the temperature. In the case of spin supercurrents, however, the Ginzburg-
Landau coherence length $\xi$ is not only a function of temperature,
but also a function of the difference between the HPD precession frequency
and the local Larmor frequency, according to Eq.(\ref{Ksi}). This
quantity can be varied experimentally
with a magnetic field gradient or position of the domain boundary. As a
result one is able to change the coherence length in the region of the orifice in
the channel and observe the change  from the canonical current-phase relation to
phase slip behavior.
This experiment made in Kapitza Institute    \cite{R1,7,72}  is schematically presented in Fig. \ref{Jos}.  The orifice, of diameter 0.48 mm, was placed in the
central part of the channel. The current-phase characteristics, observed in this
experiment are represented  for different positions of the domain
boundary related to the orifice. One can easily see that the current in Fig. (a)
corresponds to the canonical current-phase relation, which transforms  to the nonlinear
relation in Fig. (b) and then  to a phase slip phenomenon in Fig. (c).

 In the insertion the modification of the channel profile and screen for the observation of Josephson phenomena is shown.

The first attempt to describe theoretically the spin supercurrent Josephson phenomenon
 was made in \cite{Mark}. In spite of some
difficulties in presenting a simple mathematical model of the spin
supercurrent in an orifice, his calculations have a qualitative agreement
with the observed phenomena.

\section{Magnon BEC states}

The formation of magnon BEC states was confirmed by a many observation. First of all it is the observation of  Nambu-Goldstone (NG) mode of magnons condensate \cite{Goldstoun1,Goldstoun2}.
This modes are  the magnetic analog of second sound in superfluid $^4$He. It is very important achievement to support the BEC state permanently. It allow to perform the steady state experiments with a two BECs  connected by a channel. The spin supercurrent \cite{SpinCurrent}, phase slippage \cite{Slippage}, Josephson effect \cite{7,72}, spin vortex\cite{vortex,vortex2} and other supermagnonic quantum phenomena was observed.

There was discovered the other types of magnon BEC states in $^3$He-B; the self-trapped BEC state, named Q-ball \cite{Qball}, the state at a global minimum of dipole-dipole energy, named HPD2 \cite{HPD2} and the state with partial magnetization \cite{Dm2,semi}. The magnon BEC was suggested in superfluid $^3$He-A 
\cite{HeAteor} and it was observed  by pulse \cite{He-A-puls} and CW \cite{He-A,RelaxA} methods. Recently the magnon BEC and spin superfluidity was observed in a new antiferromagnetic superfluid state - $^3$He-P \cite{He-P}.

There are no any specific properties of antiferromagnetic superfluid states of $^3$He, which give advantage for magnon BEC formation, except the very small Gilbert damping factor of magnons, which can be as low as $10^{-8}$. Indeed it should be possible to found magnon BEC phenomena in other solid magnetics, as was predicted in \cite{BunAFM}. Particularly it was very interesting to search the magnon BEC in systems with coupled nuclear-electron precession, which properties are very similar to $^3$HE-A in aerogel \cite{BunDum}.  We have successfully found the formation of  magnon BEC in antiferromagnets $MnCO_3$ and $CsMnF_3$ at $1.5^o$K temperature by CW \cite{SNBEC} and pulsed \cite{SNBECPulse} NMR. The observation was done on quasinuclear branch of precession, which characterized by  Gilbert damping factor of about $10^{-5}$ and the repulsive interaction between magnons. 
The new techniques of magnon BEC formation was developed:the non-resonance excitation of magnon BEC 
\cite{nonres} and the switch off RF field method  \cite{noneq2}. The magnon BEC was observed even in very inhomogeneous conditions \cite{noneq}. 

A new breakthrough in research has taken place in YIG. The  BEC state of  magnons with wave vector k = 0 was observed in normally magnetized Yttrium Ferrite Garnet thin film at room temperature \cite{YIG}. This magnon BEC state differed from observed early magnon BEC state in YIG film magnetized tangentially  \cite{YIGGer}, where magnons with non-zero k were condensed.

\section{Acknowledgments} 

 The authors wish to thank G. E. Volovik, V. P. Mineev,
V. Lvov and O. A. Serga for helpful comments. This
work was financially supported by the Russian Science
Foundation (grant RSF 16-12-10359).


\begin{thebibliography}{9}

\bibitem{A} 
Kouki Nakata, Pascal Simon \& Daniel Loss  ``Spin Currents and Magnon Dynamics in Insulating Magnets''
J. Phys. D: Appl. Phys. 50, 114004 (2017)

\bibitem{B} O. Gomonay, V. Baltz, A. Brataas \& Y. Tserkovnyak ``Antiferromagnetic spin textures and dynamics''
 Nature Physics 14(3)  DOI: 10.1038/s41567-018-0049-4 (2018).

\bibitem{C} Rezende, S. M., Rodríguez-Suárez, R. L. \& Azevedo, A. ``Diffusive magnonic spin transport in antiferromagnetic insulators'' Phys. Rev. B 93,
054412 (2016).


\bibitem{Einstein} Einstein, A. "Quantentheorie des einatomigen idealen Gases. Part I". {\it Sber. Preuss.
Akad. Wiss.} {\bf 22}, 261–267 (1924); "Quantentheorie des einatomigen idealen
Gases. Part II". {\it Sber. Preuss. Akad. Wiss.} {\bf 1,} 3-14 (1925).

\bibitem{critdens} Gazizulin, R. R., Bunkov, Yu. M., Safonov, V. L. “Critical parameters of nuclear magnon Bose–Einstein condensation in systems with dynamical frequency shift”
{\it JETP Lett.}, {\bf 102}, 876 – 880, (2015).


\bibitem{Bunkov2018a} Bunkov, Yu. M., Safonov, V. L. "Magnon condensation and spin superfluidity". {\it Journal of Magnetism and Magnetic Materials} {\bf 452,} 30–34 (2018).

\bibitem{Volovik1987}
G.E. Volovik,
Linear momentum in ferromagnets,
J. Phys. C {\bf 20}, L83-87 (1987).

\bibitem{MagBEC} Bunkov, Yu. M. \& Volovik, G. E. ``Bose-Einstein Condensation of Magnons in Superfluid $^{3}$He". {\it J. of Low Temp. Phys.} {\bf 150,} 135-144 (2008).

\bibitem{HPD} Borovik-Romanov, A. S., Bunkov, Yu. M., Dmitriev, V. V. \& Mukharskii, Yu. M.
``Long-lived induction signal in superfluid $^{3}$He-B". {\it JETP Lett.} {\bf 40,} 1033-1037 (1984).

\bibitem{HPDT} Fomin, I. A. ``Long-lived induction signal and spatially nonuniform spin precession in $^{3}$He-B". {\it JETP Lett.} {\bf 40,} 1037-1040 (1984).



\bibitem{2} Kagan, Y. \&  Manakova, L. A. ``Condensation of phonons in an ultracold Bose gas". {\it Phys. Lett. A} {\bf 361,} 401 (2007).

\bibitem{3} Butov L. V. {\it et al.} ``Stimulated Scattering of Indirect Excitons in Coupled Quantum Wells: Signature of a Degenerate Bose-Gas of Excitons". {\it Phys. Rev. Lett.} {\bf 86,} 5608 (2001).

\bibitem{4} Kasprzak, J. {\it et al.} ``Bose-Einstein condensation of exciton polaritons". {\it Nature (London)} {\bf 443}, 409 (2006).

\bibitem{5} Klaers, J., Schmitt J., Vewinger, F., \&  Weitz,M. ``Bose-Einstein condensation of photons in an optical microcavity" {\it Nature (London)} {\bf 468,} 545 (2010).

\bibitem{6} Melnikovsky, L. A. ``Bose-Einstein condensation of rotons". {\it Phys. Rev. B} {\bf 84,} 024525 (2011).

\bibitem{Tcrystal} Volovik, G. E. ``On the broken time translation symmetry in macroscopic systems: Precessing states and off-diagonal long-range order". {\it JETP Lett.} {\bf 98,} 491-495 (2013).

\bibitem{Goldstoun1} Bunkov, Yu. M., Dmitriev, V. V. \& Mukharskii, Yu. M. ``Torsional vibrations of a domain with uniform magnetization precession in $^{3}$He-B". {\it JETP Lett.} {\bf 43,} 131-134 (1986).

\bibitem{Goldstoun2} Bunkov, Yu. M., Dmitriev, V. V. \& Mukharskii, Yu. M. ``Low frequency oscillations of the homogeneously precessing domain in $^{3}$He-B". {\it Physica B} {\bf 178,} 196-201 (1992).

\bibitem{Goldstoun3} Yu. M. Bunkov,  A. V. Klochkov, T. R. Safin  K. R. Safiullin, M. S. Tagirov
``Goldstone Mode of a Magnon Bose-Einstein Condensate in MnCO$_3$''
JETP Lett., 106,  677-681  (2017)



\bibitem{JETPBEC} Borovik-Romanov, A. S., Bunkov, Yu. M., Dmitriev, V. V., Mukharskii, Yu. M. \& Flachbart, K. "Experimental study of separation of magnetization precession in $^{3}$He-B into two magnetic domains".{\it JETP} {\bf 61,} 1199-1206 (1985).

\bibitem{Snoke} Snoke, D. "Coherent questions". {\it Nature} {\bf 443,} 403-404 (2006).

\bibitem{mBEC} Bunkov, Yu. M. "Magnon BEC Versus Atomic BEC". {\it J. Low Temp. Phys.} {\bf 185,} 399-408 (2016).




\bibitem{CWBEC} Borovik-Romanov, A.S.  {\it et al.}
"Distinctive Features of a CW NMR in 3He-B due to a Spin  Supercurrent", {\it JETP} {\bf 69,} 542 (1989).

\bibitem{Rev1} Bunkov, Yu. M. "Spin superfluidity and coherent spin precession". {\it J. Phys.: Condens. Matter} {\bf 21,} 164201 (2009).

\bibitem{Rev2} Bunkov, Yu. M. \& Volovik, G. E. "Magnon Bose–Einstein condensation and spin superfluidity". {\it J. Phys.: Condens. Matter} {\bf 22,} 164210 (2010).

\bibitem{62} A.S.Borovik-Romanov, Yu.M.Bunkov, V.V.Dmitriev, Yu.M.Mukharskiy and D.A.Sergatskov,
Investigation of spin supercurrent in $^3$He-B,
Phys. Rev. Lett. {\bf 62}, 1631--1634 (1989).

\bibitem{R1} Yu.M. Bunkov,
Spin supercurrent and novel properties of NMR in superfluid $^3$He,
 in:  Prog. Low Temp. Phys. Vol XIV, p. 69, ed. W. Halperin, Elsevier, Amsterdam (1995).



\bibitem{63}
Yu.M.Bunkov, V.V.Dmitriev,  Yu.M.Mukharskiy and G.K.Tvalashvily,
Superfluid spin current in  a channel parallel to the magnetic field,
Sov. Phys. JETP {\bf 67}, 300 (1988).


\bibitem{Einzel} Einzel, O., 1981, Physica 108B, 1143.

\bibitem{Cat}   Bunkov, Yu. M. {\it et al.,}   Catastrophic Relaxation in 3He-B at 0.4 T$_c$  {\it Europhysics Letters},  {\bf 8}, 645 (1989).



\bibitem{Kopnin} Ivlev, B.1. and N.B. Kopnin, 1984, {\it Sov. Phys. Uspekhy} {\bf 27}, 206.

\bibitem{Varoquaux} Varoquaux, E., O. Avenel. G. lhas and R. Salomelin. 1992, {\it Physica B} 178, 309.





\bibitem{Fominslip} Fomin, I.A., 1988, {\it Sov. Phys. JETP} {\bf 67}, 1148.



\bibitem{VollhardtWolfle1990}
Vollhardt  D. and W\"olfle P. (1990).
{\it The superfluid phases of helium 3},  Taylor and Francis, London.

\bibitem{Sauls} Sauls, J. A., Bunkov, Yu. M., Collin, E., Godfrin, H. \& Sharma, P. "Magnetization and spin diffusion of liquid $^{3}$He in aerogel". {\it Phys. Rev. B} {\bf 72,} 024507 (2005).


\bibitem{Joseph}
B.D. Josephson,
Possible new effects in superconducting tunnelling,
{\it Phys. Lett.} {\bf 1}, 251--253 (1962).

 \bibitem{7}
 A.S. Borovik-Romanov, Yu.M. Bunkov, A. de Waard,
V.V. Dmitriev, V. Makrotsieva, Yu.M. Mukharskiy, D.A. Sergatskov,
Observation of a spin supercurrent analog of the Josephson  effect,
{\it JETP Lett.} \textbf{47}, 478--482 (1988).

\bibitem{72}
A.S.Borovik-Romanov, Yu.M.Bunkov, V.V.Dmitriev, Yu.M.Mukharskiy and D.A.Sergatskov,
Josephson effect in spin supercurrent in $^3$He-B,
AIP Conf. Proc. {\bf 194}, 27 (1989).

\bibitem{Mark}
A.V. Markelov,
Josephson effect on a spin current,
{\it JETP} {\bf 67}, 520--523 (1988).



\bibitem{SpinCurrent} Borovik-Romanov, A. S., Bunkov, Yu. M., Dmitriev, V. V., Mukharskiy, Yu. M. \& Sergatskov, D. A. "Investigation of Spin Supercurrent in $^{3}$He-B". {\it Phys. Rev. Lett.} {\bf 62,} 1631-1634 (1989).

\bibitem{Slippage} Borovik-Romanov, A. S., Bunkov, Yu. M., Dmitriev, V. V. \& Mukharskii, Yu. M.
"Observation of phase slippage during the flow of a superfluid spin current in $^{3}$He-B". {\it JETP Lett.} {\bf 45,} 124-128 (1987).



\bibitem{vortex} Borovik-Romanov, A. S.  {\it et al.}
"Observation of Vortex-like Spin Supercurrent in $^{3}$He-B". {\it Physica B} {\bf 165,} 649-650 (1990).

\bibitem{vortex2} Bunkov, Yu. M. \& Volovik, G. E. "Spin vortex in magnon BEC of superfluid $^{3}$He-B". {\it Physica C} {\bf 468,} 609-612 (2008).

\bibitem{Qball} Autti, S. {\it et al.} "Self-Trapping of Magnon Bose-Einstein Condensates in the Ground State and on  Excited Levels: From Harmonic to Box Confinement". {\it Phys. Rev. Lett.} {\bf 108,} 145303 (2012).

\bibitem{HPD2} Elbs, J.  {\it et al.}
"Strong Orientational Effect of Stretched Aerogel on the $^{3}$He Order Parameter". {\it Phys. Rev. Lett.} {\bf 100,} 215304 (2008).



\bibitem{Dm2} Dmitriev, V. V. {\it et al.} "Stable Spin Precession at One Half of Equilibrium Magnetization in Superfluid $^3$He-B" {\it Phys. Rev. Lett.},
{\bf 78}, 86-89 (1997).

\bibitem{semi} Bunkov, Yu. M. {\it et al.},
Semi-superfluidity of 3He in Aerogel
{\it Phys. Rev. Lett.} {\bf 85}, 3456 (2000).


\bibitem{HeAteor} Bunkov, Yu. M. \& Volovik, G. E. "On the possibility of the Homogeneously Precessing Domain in Bulk $^{3}$He-A". {\it Europhys. Lett.} {\bf 21,} 837 (1993).

\bibitem{He-A-puls} Hunger, P., Bunkov, Yu. M., Collin, E. \& Godfrin, H. ``Evidence for Magnon BEC in Superfluid $^{3}$He-A". {\it J. of Low Temp. Phys} {\bf 158,} 129–134 (2010).

\bibitem{He-A} Sato, T. {\it et al.} ``Coherent Precession of Magnetization in the Superfluid $^{3}$He A-Phase". {\it Phys. Rev. Lett.} {\bf 101,} 055301 (2008).

\bibitem{RelaxA}  Matsubara, A. {\it et al.}  “Coherent precession of magnetization in superfluid 3He A-phase in aerogel”
{\it J. Phys.: Conf. Ser.} {\bf 150} 032052 (2009).

\bibitem{He-P} Autti, S. {\it et al.} "Bose-Einstein condensation of magnons and spin superfluidity in the polar phase of $^{3}$He" {\it Phys. Rev. Lett.},
{\bf 121}, 025303 (2018).

\bibitem{BunAFM}  Bunkov, Yu.M. ``Spin superfluidity and magnons Bose–Einstein condensation''  {\it Physics Uspekhi}, {\bf 53,} 848 (2010). 

\bibitem{BunDum} Bunkov,Yu.M. \& Dumesh,B.S.
  "Dynamic Properties of Pulsed NMR at Easy Plane Antiferromagnets
   with Large Pulling"  {\it Sov. Phys. JETPh} {\bf 41,} 576 (1975).


\bibitem{SNBEC} Bunkov, Yu. M. {\it et al.} "High-T$_{c}$ Spin Superfluidity in Antiferromagnets". {\it Phys. Rev. Lett.} {\bf 108,} 177002 (2012).

\bibitem{SNBECPulse} Tagirov, M. S. {\it et al.} "Magnon BEC in Antiferromagnets with Suhl-Nakamura Interaction". {\it J. Low Temp. Phys.} {\bf 175,} 167-176 (2014).



\bibitem{nonres} Bunkov, Yu. M. {\it et al.} JETP Lett. {\bf 109}, 43 (2019).

\bibitem{noneq} Bunkov, Yu. M. \& Tagirov, M. S.
"Magnon Bose-Einstein condensation at inhomogeneous conditions". {\it Journal of Physics: Conference Series} {\bf 478,} 012004 (2013).

\bibitem{noneq2} Bunkov, Yu. M. "The magnon BEC observation by switch off method". {\it Low Temp. Phys} {\bf 43,} 1158-1160 (2017).

\bibitem{YIG} Bunkov, Yu. M. {\it et al.} ``Conventional magnon BEC in YIG film'' http://arxiv.org/abs/1810.08051v2 (2018)

\bibitem{YIGGer} Serga, A. A., Chumak, A. V. \& Hillebrands, B. ”YIG
magnonics”, {\it J. Phys. D: Appl. Phys} {\bf 43}, 264002
(2010).

\end{thebibliography}
\end{document}